\documentclass
[10pt,tightenlines,showpacs,letterpaper,twocolumn,runinaddress,prd]{revtex4}
\usepackage{amsfonts}
\usepackage{amsmath}
\usepackage{amssymb}
\usepackage{graphicx}
\begin{document}
\title{Gaps below strange star crusts}
\author{Morten Stejner and Jes Madsen}
\affiliation{Department of Physics and Astronomy, University of Aarhus, DK-8000 \AA rhus C, Denmark}

\begin{abstract}
The gap caused by a strong electric field
between the quark surface and nuclear crust of a strange star is
studied in an improved model
including gravity and pressure as well as electrostatic
forces. The transition from gap to crust is followed in detail.
The properties of the gap are investigated for a wide range of parameters
assuming both color-flavor locked and non color-flavor locked strange
star cores. The maximally allowed crust density is generally
lower than that of neutron drip.
Finite temperature is shown to increase the gap width, but the effect
is significant only at extreme temperatures.
Analytical approximations are derived and shown to provide useful fits
to the numerical results.
\end{abstract}
\date{October 31, 2005}
\maketitle

\section{Introduction}
Strange stars are stars made of absolutely stable quark matter
\cite{Baym:1976yu,Witten:1984rs,Haensel:1986qb,Alcock:1986hz}.  Their
existence (i.e., the stability of strange quark matter) depends on
poorly constrained strong interaction properties, and remains to be
decided by observation or experiment (see
\cite{Madsen:1998uh,Weber:2004kj} for reviews). If strange stars are
stable, they contain roughly equal numbers of up, down, and strange
quarks, but due to the higher mass of the s-quark, they will normally
contain a slight deficit of strange quarks with negative charge, and
thus have an overall positive quark charge to be compensated by
electrons. Even color-flavor locked quark matter, which is
electrically neutral in bulk \cite{Rajagopal:2000ff}, has an overall
positive quark charge due to surface effects
\cite{Madsen:2000kb,Madsen:2001fu,Usov:2004iz}.  The quark surface of
a strange star is very sharp (the density drops from above nuclear
matter density to zero within a few fm), and since the electrostatic
force is weaker than the strong force, some of the electrons necessary
to create an overall charge neutral object will form a thin atmosphere
with a huge electric field (up to $10^{18}$ V/cm) outside the quark
phase. This field is capable of sustaining a nuclear matter crust
decoupled from the quark phase by an electron-filled gap.  The maximum
mass of such a crust is approximately $2\times 10^{-5} M_\odot$,
corresponding to the situation where the inner boundary of the crust
reaches the neutron drip density, $4\times 10^{11}\; {\rm g\;
cm}^{-3}$, where neutrons drip out of crust nuclei and get dissolved
in the quark phase. Smaller crust mass limits would occur if the gap
is sufficiently narrow to allow direct contact with the crust or if
the rate of quantum tunneling of nuclei into the quark core of the
star is large. The crust mass and moment of inertia as well as the
coupling between core and crust play important roles in the
understanding of strange star properties, and therefore the properties
of the gap are very important.  The width of this gap is the main
topic of the present investigation.

The properties of the electron atmosphere and the impact on a nuclear
matter crust have been studied by many authors, including
\cite{Alcock:1986hz,Kettner:1994zs,Huang:1997,Xu:1999up,Cheng:2003hv,
Usov:2004iz,Usov:2004kj,Jaikumar:2005,Zdunik:2002}.  These studies
have involved solutions to the Poisson equation for the electrostatic
potential with the boundary condition of electric neutrality deep
within the strange star, as well as a condition for the potential at
large distances. For bare strange stars, i.e., stars without a nuclear
crust, the potential goes to zero at infinity. For strange stars
accreting nuclear matter on the surface, the condition has been taken
to be a matching of the electric potential from the electrons to the
value of the potential required in the bulk of the nuclear matter
crust (typically several MeV, depending on the crust density). This
leads to a typical gap size (distance from quark surface to the inner
surface of the nuclear crust) of $10^{2}-10^{3}$ fm, comparable to the
distance over which the potential drops by a factor of a few.

The approaches just described do not account however for the detailed
balance between electrical and gravitational forces and pressure in the
transition from gap to crust. The potential and its derivative are
taken to be continuous across the crust boundary while gravitational
forces and pressure in the crust are assumed insignificant
(though see \cite{Huang:1997} for an approximate inclusion of gravity). 
In this work we expand earlier treatments, investigate the effects of including
the Newtonian gravitational field, and find the detailed structure of the gap
and the transition to the crust. This leads to more narrow gaps than
previously found and thus constrains the density at the base of the
crust below the neutron drip density. As the temperature increases,
the additional thermal reservoir of electrons initially widens the gap
then narrows it, but even for temperatures as high as $10^7$ K in the
gap region this effect is small and the width remains constant up to
temperatures in the range $10^8-10^{10}$ K depending on the crust
density.

\section{Equilibrium of the crust}
The effective chemical potential of species $i$, 
$\mu_{i}^{\text{eff}}$, is defined as the change in energy for
a unit change in number density of species $i$, while the volume,
entropy, and other number densities are kept constant.
In a steady state assuming the crust
to be isothermal the effective chemical potentials of electrons and
nuclei should be constant to avoid migration. They must therefore be equal to
their values at the top of the crust, at radius $r=R$, where the electric
potential is zero and the chemical potentials are equal to the particle
masses. Therefore we
take the effective chemical potentials for electrons and nuclei to be
\begin{equation}
\label{mueff}
\begin{split}
\mu_e^{\text{eff}}=&\mu_e(r)-e\phi_e(r)+m_e\phi_g(r)=m_e+m_e\phi_g(R) \\
            &\mu_e(r)=m_e+m_e[\phi_g(R)-\phi_g(r)]+e\phi_e(r)\\
\mu_N^{\text{eff}}=&\mu_N(r)+Ze\phi_e(r)+m_N\phi_g(r)=m_N+m_N\phi_g(R)\\
            &\mu_N(r)=m_N+m_N[\phi_g(R)-\phi_g(r)]-Ze\phi_e(r).
\end{split}
\end{equation}
For simplicity we describe electrons as well as nuclei as Fermi gases
characterized by thermodynamic chemical potentials $\mu_e$ and $\mu_N$.
$m_e$ is
the electron mass, $m_N$ is the mass of the dominant nucleus at the
base of the crust, $Z$ is its charge, $\phi_e(r)$ is the electric
potential and $\phi_g(r)$ is the gravitational potential for which we
will use the Newtonian value to get tractable expressions. $e\phi_e$
is typically of the order of 20 MeV at the strange core surface,
$r=R_S$, and $\phi_g(R_S)=-GM_S/R_S\sim -0.2$, where $M_S$ is
the mass of the strange core.

We note at this point that keeping the effective chemical potentials
constant corresponds to hydrostatic equilibrium which may be seen
explicitly by taking the gradient of the above expressions and using
the zero temperature identity $dP_i=n_id\mu_i$, where $P_i$ is the
pressure and $n_i$ the number density of the $i$'th component.
\begin{align*}
&\frac{dP_e}{dr}-n_e\frac{d(e\phi_e)}{dr}+n_em_e\frac{d\phi_g}{dr}=0\\
&\frac{dP_N}{dr}+Zn_N\frac{d(e\phi_e)}{dr}+n_Nm_N\frac{d\phi_g}{dr}=0
\end{align*}    
The total pressure is $P=P_e+P_N$ so the sum then gives the usual
force balance
\begin{align}
\frac{dP}{dr}+n_q\frac{d(e\phi_e)}{dr}+\rho\frac{d\phi_g}{dr}=0\label{forcebalance}
\end{align}
with $n_q=Zn_N-n_e$ the charge density and $\rho=m_en_e+m_Nn_N$ the
mass density. We show later that Eq.~\eqref{forcebalance} is convenient 
when approximating thin crusts.

To solve for the chemical potentials we take the Laplacian of
Eq.~\eqref{mueff} and use Poisson's equation in Newtonian
gravity to get a system of coupled differential equations in the
chemical potentials and their derivatives. With the understanding that
the core at $r<R_S$ is comprised of quarks and electrons while only
electrons are present in the gap at $R_S<r<R_C$, and that nuclei do
not appear until $\mu_N>m_N$ at $r>R_C$ this gives in a compact notation
\begin{align}
  \begin{split}
    \nabla^2\mu_e(r)=&\frac{d^2\mu_e}{dr^2}+\frac{2}{r}\frac{d\mu_e}{dr}\\
    =&\nabla^2[e\phi_e(r)-m_e\phi_g(r)]\\
    =&-4\pi e^2(n_q^+ + Zn_N - n_e)\\&-4\pi Gm_e(\rho_q+m_Nn_N+m_en_e)\label{mue}
  \end{split}\\
  \begin{split}
    \nabla^2\mu_N(r)=&\frac{d^2\mu_N}{dr^2}+\frac{2}{r}\frac{d\mu_N}{dr}\\
    =&\nabla^2[-Ze\phi_e(r)-m_N\phi_g(r)]\\
    =&4\pi e^2Z(n_q^+ + Zn_N - n_e)\\&-4\pi Gm_N(\rho_q+m_Nn_N+m_en_e)\label{muN}
  \end{split}
\end{align}   
Here $\rho_q$ is the approximately constant mass density of the quark
matter and $n_q^+$ is its positive charge density -- we assume that
the quark distribution is not significantly affected by the crust (see
however the recent work by Jaikumar et al.~\cite{Jaikumar:2005}). In
these equations we have used rest mass times number density for the matter
density in Poissons equation. This is a very rough approximation for
the electrons as they are relativistic except for very thin crusts,
but in all the calculations below we will neglect the electron
contribution to the density completely by taking
$m_e=0$. The densities of electrons and nuclei are found from
\begin{align}
n_i=\int d^3p_i \frac{g_i}{h^3}f(p_i)\;,\qquad f(p_i)=\frac{1}{e^{((E_i-\mu_i)/T)}+1}
\end{align}
where $p$ is the particle momentum, $g_i$ the statistical weight,
$E_i=\sqrt{p_i^2-m_i^2}$ the energy and $f(p)$ the Fermi-Dirac
distribution. For a cold Fermi gas with $g_i=2$ this gives 
\begin{align}
n_i=\frac{(\mu_i^2-m_i^2)^{3/2}}{3\pi^2}\label{n}\; .
\end{align}
while for finite temperatures there will be additional thermal electrons and
working in the approximation $m_e=0$ the net electron density (electrons
minus positrons) will be
\begin{align}
n_e=\frac{\mu_e^3}{3\pi^2}+\frac{\mu_eT^2}{3} \label{ne}
\end{align}        
Equations \eqref{mue} and \eqref{muN} can thus be transformed into four first
order coupled differential equations in the four unknowns
$\mu_e,\mu_N,\frac{d\mu_e}{dr},\frac{d\mu_N}{dr}$. Given appropriate
boundary conditions at some point -- usually either in the charge
neutral bulk of the strange star core or at the core surface -- the
system can be solved to find the structure of the gap and the
transition to the crust. In principle we could extend the model to
include the entire crust, but it should be noted that we have not
included the interactions between nuclei (apart from the mean electric
potential and gravity) and that the global
properties of the crust should be described using relativistic gravity
and a more advanced equation of state -- such as the BPS equation of
state. Here we
restrict ourselves to modeling the gap and the lower crust until the
point where charge neutrality is achieved. The density at this point
can then be taken as the inner crust density used to parametrize the
crust models in e.g. \cite{Zdunik:2002}.

\section{Approximative models} \label{apprmodel}
Before we get into the detailed numerical models it is worth noticing
that a number of illustrative approximate models can be calculated
analytically, and that as it turns out these models reproduce the
behavior of the system accurately. 

First of all it should be noted that we can get a very useful
relation from Eq.~\eqref{mueff} by taking the sum
\begin{align}
\mu_N(r)+Z\mu_e(r)=(m_N+Zm_e)[1+\phi_g(R)-\phi_g(r)]\label{sum}
\end{align}
and its derivative
\begin{align}
\frac{d\mu_N(r)}{dr}+Z\frac{d\mu_e(r)}{dr}=-(m_N+Zm_e)\frac{d\phi_g(r)}{dr}.
\label{diffsum}
\end{align}
From this and equations (\ref{mue}, \ref{muN}) we can get a
qualitative idea of the form of the solutions. For example we know
that just above the strange core surface the electric potential should
be large and positive and its gradient should be large and negative
while the gravitational potential and its gradient can be approximated
by their Newtonian values. Only electrons are present at this point so
the charge density is negative, and $\mu_e$ will be large and
decreasing rapidly but with a positive curvature. Since the right hand
side of equation \eqref{sum} is almost constant across the gap
$\mu_N$ must increase as $\mu_e$ decreases. Thus we also see that
$\mu_e$ must decrease monotonically since $\mu_e$ and $\mu_N$ would
diverge in opposite directions if $\mu_e$ started increasing while the
charge density was still negative and $\mu_e(r)$ was curving upwards. Once
$\mu_N>m_N$ we enter the crust and at some point achieve charge
neutrality. $\mu_N(r)$ curves downwards only if the charge density is
negative and since we expect the crust to be charge neutral in bulk $\mu_N$
must start decreasing along with $\mu_e$ at this point for the crust
to remain charge neutral. The right hand side of equation
\eqref{diffsum} is of the order of $\sim -10^{-15} \text{ MeV/fm}$ so
for both $\frac{d\mu_e}{dr}$ and $\frac{d\mu_N}{dr}$ to be
negative they must both be numerically smaller than $10^{-15}\text{
MeV/fm}$. Since $\frac{d\mu_e}{dr}$ will generally be of the order
of $0.1 \text{ MeV/fm}$ close to the surface of the strange star core this
places high demands on the numerical procedure as it must be accurate
across these many orders of magnitude. This difficulty stems from the
fact that we are trying to model the transition from the gap where the
strong electrical force is dominant to a charge neutral crust held down
by the much weaker gravitational force -- a very stiff set of equations.

To get something more quantitative than these heuristic arguments we
will use the approximation $m_e=0$. This allows us to solve for
$\mu_e$ in the case of a pure electron atmosphere with no crust. This
can then be used to find approximate expressions for the gap
width. For more realistic solutions it should be taken into account that some of
the electrons from the crust will spill into the gap shielding the
strange star core surface charge and affecting the gap width. We study the
full solutions numerically in a later section, but these are much
easier to understand (and find) with the insight from 
the qualitative models below.

\subsection{The pure electron atmosphere around a bare strange star}
\begin{figure}[!htbp]
\centering
\includegraphics[width=8.5cm]{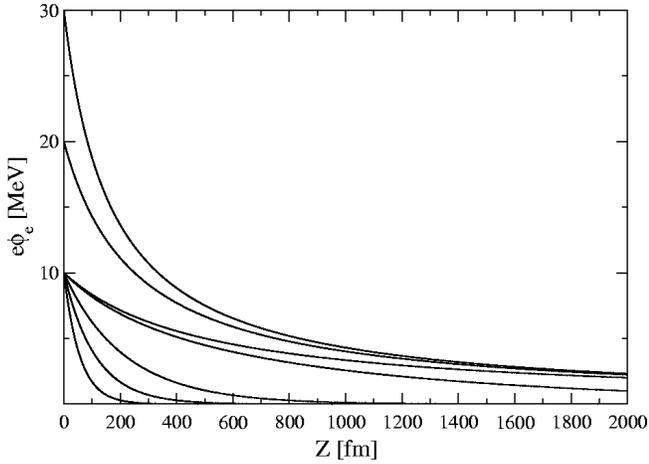}
\caption{Electric potential above the surface of a
  bare strange star. From the top $e\phi_e(R_S)=30, 20, 10 \text{
  MeV}$. We show the variation with temperature for $e\phi_e(R_S)=10
  \text{ MeV}$ so from the top $T=0,1,5,10,20 \text{ MeV}$ for these.}
\label{potential}
\end{figure}

For a pure electron atmosphere with $m_e=0$ we have
$\mu_e(r)=e\phi_e(r)$ and in general
\begin{align}
n_e=\frac{(e\phi_e)^3}{3\pi^2}+\frac{(e\phi_e)T^2}{3}\;.
\end{align}
Following \cite{Cheng:2003hv,Usov:2004kj} we define $y=d(e\phi_e)/dz$ where $z
$ is the height above the strange core surface. Assuming a flat strange
core surface Eq.~\eqref{mue} then takes the form
\begin{align}
\frac{d(y^2)}{d(e\phi_e)}=4C^{-2}((e\phi_e)^3+\pi^2T^2(e\phi_e))
\end{align}  
where $C=\sqrt{3\pi/2}/e=5013 \text{ MeV fm}$. This may be solved
directly for the electric field and potential using the fact that both
should go to zero at large $z$:
\begin{align}
&-\frac{d(e\phi_e)}{dz}=C^{-1}((e\phi_e)^4+2\pi^2T^2(e\phi_e))^{1/2}\label{felt}\\
&z=\frac{C}{\sqrt{2}\pi T}\left[\sinh^{-1}\left(\frac{\sqrt{2}\pi
    T}{e\phi_e(r)}\right)-\sinh^{-1}\left(\frac{\sqrt{2}\pi T}{e\phi_e(R_S)}\right)\right]\label{z}
\end{align} 
where we give $z(e\phi)$ as this is what we will need later. It may be
noted that in the limit $T\rightarrow 0$ equation \eqref{z} reduces to
the expression found by Alcock et al.\ in \cite{Alcock:1986hz} since
$\sinh^{-1}x\sim x$ for $x\rightarrow 0$
\begin{align}
e\phi_e(r)=\frac{C}{z+C/e\phi_e(R_S)} \text{ for } T=0\; .\label{coldpotential}
\end{align}
A plot of this behavior can be found in Fig.~\ref{potential}.
Note that we have included the case $e\phi_e(R_S)=30~\text{MeV}$. 
In the models below this just allows a crust at neutron drip density to remain
out of direct contact with the strange core surface making it an illustrative
case, though the choice is difficult to realize for non color-flavor
locked quark matter equations of state \cite{Fu:2003}.

\subsection{Thin crusts and test particles}
Sufficiently thin crusts should not be expected to change the electric
potential or the distribution of electrons in the gap from that of a
pure electron atmosphere. We can then use the solution found in the
previous subsection and demand that equation \eqref{forcebalance} is
obeyed at the crust boundary. Since we are working in the
approximation $m_e=0$ the electronic part of the force balance is
simply $\frac{dP_e}{dr}-n_e\frac{d(e\phi_e)}{dr}=0$ which is trivially
fulfilled since $\mu_e=e\phi_e$ and $dP_e=n_ed\mu_e$. For a thin crust
we should be able to ignore the pressure gradient from the nuclei and the
remainder of the force balance equation involves only the equilibrium
between electrostatic repulsion and gravitational attraction of
positively charged nuclei:
\begin{align}
Zn_N\frac{d(e\phi_e)}{dr}+\frac{GM_Sn_Nm_N}{r^2}=0\label{simplebalance}
\end{align}  
The number density of nuclei then cancels and the equation simply
expresses that each individual nucleus feels a balance between the
gravitational attraction from the strange star core and the
electrostatic repulsion from the net positive charge below it as
described by $d(e\phi_e)/dr$. This is the same equation as would be
obtained by balancing a single positively charged test particle above
the strange star surface. In the case of a cold atmosphere we use the
derivative of equation \eqref{coldpotential} for the electric field which
leads to the relation
\begin{align}
-\frac{C}{(z_{gap}+C/e\phi_e(R_S))^2}=-\frac{GM_Sm_N}{ZR_C^2}\; ,
\end{align}
or $z_{gap}\equiv R_C-R_S\simeq R_S(CZ/GM_Sm_N)^{1/2}\simeq 
10^{10}~\text{fm}$. The gap width of very thin crusts is therefore almost
macroscopic in size as might be expected since the lower layer does not
have to support the bulk of the crust. If we assume that equation
\eqref{simplebalance} is obeyed identically throughout the entire
crust as it should be if the pressure gradient from the nuclei remains
negligible we can solve for the potential
\begin{align}
e\phi_e(r>R_C)=\frac{GMm_N}{Zr}+K\; ,
\end{align}
where $K$ is a constant. This potential is a solution to Poisson's
equation without a source term, 
\begin{align}
\nabla^2(e\phi_e)=\frac{d^2(e\phi_e)}{dr^2}+\frac{2}{r}\frac{d(e\phi_e)}{dr}=0
\end{align} 
showing that the crust is locally charge neutral in this
approximation. 

\subsection{Cold charge neutral crusts}
In heavy dense crusts the pressure from the nuclei can not be ignored,
and we would need the precise density profile to solve equation
\eqref{forcebalance}. The crust would still be expected to be charge
neutral beyond some point not very far from the crust boundary however,
so by assuming charge neutrality and using equation \eqref{sum} it
is possible to estimate $\mu_e$ at the crust boundary. The gap width
can then be found assuming that the potential in the gap can be described
by the solution for a pure electron atmosphere. 

For a cold crust charge neutrality in the Fermi gas approximation
is described by the requirement
\begin{align}
\frac{(\mu_e^2-m_e^2)^{3/2}}{3\pi^2}=Z\frac{(\mu_N^2-m_N^2)^{3/2}}{3\pi^2}.
\end{align}
Solving for $\mu_e$ and using relation \eqref{sum} this gives
\begin{equation}
\begin{split}
\mu_e=&\bigg((m_N+Zm_e)[1+\phi_g(R)-\phi_g(r)]-\\
      &\Big[(m_N^2-Z^{-2/3}m_e^2)(1-Z^{-8/3})+\\
      &Z^{-8/3}\big[(m_N+Zm_e)\\
      &(1+\phi_g(R)-\phi_g(r))\big]^2\Big]^{1/2}\bigg)(Z(1-Z^{-8/3}))^{-1}.
\end{split}
\end{equation}
Taking $m_e=0$ and using $\phi_g(R)-\phi_g(r)\ll 1$ this simplifies to
\begin{equation}
\begin{split}
\mu_e=\frac{m_N}{Z}\Big[&(\phi_g(R)-\phi_g(r))\\
                        &-\frac{1}{2}(Z^{8/3}-1)[\phi_g(R)-\phi_g(r)]^2)\Big].
\end{split}
\end{equation}
Using equation \eqref{sum} again and $\rho=n_Nm_N$ this leads to a relation between the
size, mass, density and composition of the crust 
\begin{align}
&\phi_g(R)-\phi_g(r)=\Big[2\frac{\mu_N-m_N}{m_N}(Z^{8/3}-1)\Big]^{1/2}\\
&=1.9\times 10^{-4}\left(\frac{\rho}{\rho_D}\right)^{1/3}\left(\frac{56}{A}\right)^{4/3}\left(Z^{8/3}-1\right)^{1/2},\label{deltaphi}
\end{align}
where $\rho_D=4\times 10^{11}\text{ g/cm}^3$ is the neutron drip density. 
Assuming that charge
neutrality occurs shortly after the crust boundary at $R_C$, $\phi_g(r)$ will
not change much between these two points 
and we can use this to estimate $\mu_e(R_C)$. The
crust boundary is defined by $\mu_N=m_N$ and from equation \eqref{sum}
we then have $\mu_e(R_C)=\frac{m_N}{Z}(\phi_g(R)-\phi_g(R_C))$ which can be
used with the potential \eqref{coldpotential} to find the gap width
\begin{equation}
\label{approx}
\begin{split}
z_{gap}=&\frac{C}{\mu_e(R_C)}-\frac{C}{e\phi_e(R_S)}\\
=&505\left(\frac{\rho}{\rho_D}\right)^{-1/3}\left(\frac{A}{56}\right)^{1/3}Z(Z^{8/3}-1)^{-1/2}\text{
  fm}\\&-\frac{167\text{ fm}}{e\phi_e(R_S)/30 \text{MeV}}\\
=&170.5\left(\frac{\rho}{\rho_D}\right)^{-1/3}\text{fm}
-167\left(\frac{e\phi_e(R_S)}{30~\text{MeV}}\right)^{-1}\text{fm}.
\end{split}
\end{equation}
The last equality assumes a crust composed of electrons and
$^{56}\text{Fe}$ nuclei. In this approximation the gap width is
very small near neutron drip density, and for a realistic value of
$e\phi_e(R_S)\sim 20 \text{ MeV}$ it would be impossible for a crust
to reach neutron drip density while remaining out of contact with the
strange star core. If the more stringent limit of $z_{gap}>200 \text{
fm}$ from \cite{Alcock:1986hz} for the crust to remain secure against strong
interactions with the strange star core is used, the maximum crust
density is $\sim 0.1\rho_D$. We will see later that this simple
relation describes the outcome of the numerical
calculations very well.

\subsection{Charge neutral crust at finite temperatures}
At finite temperatures the electron density is given by equation
\eqref{ne}, and we will approximate the electric potential in the gap
by equation \eqref{z}. As before the crust boundary is given by
$\mu_N(R_C)=m_N$ so $\mu_e(R_C)=e\phi_e=\frac{m_N}{Z}\Delta \phi_g$ by
equation \eqref{sum} with $\Delta \phi_g=\phi_g(R)-\phi_g(R_C)$, and
we will again assume that charge neutrality occurs shortly after the
crust boundary so we can solve $n_e=Zn_N$ for $\Delta \phi_g$. Because
of the extra electrons this can not be done analytically since the
equation for $\Delta \phi_g$ now becomes
\begin{equation}
\begin{split}
0=&Z(\mu_N^2-m_N^2)^{3/2}-(m_N[1+\Delta\phi_g]-\mu_N)^3Z^{-3}\\
&-3\pi^2(m_N[1+\Delta\phi_g]-\mu_N)T^2/3Z \label{Deltaphi1}
\end{split}
\end{equation}
with
\begin{equation}
\mu_N=((3\pi^3\rho/m_N)^{2/3}+m_N^2)^{1/2} \; .
\end{equation}
These equations can however be solved numerically for
$\Delta\phi_g(\rho,T)$ which gives $\mu_e(R_C)$ and $z_{gap}$ as shown
in Fig.~\ref{artikeltemp}. Qualitatively we note that changes
from the cold case are negligible until temperatures around 1 MeV or
$10^{10} \text{ K}$ are reached -- that is until the temperature is
comparable to $\mu_e$ in the bulk of the crust. For lower densities the
effect occurs at somewhat lower temperatures, but by then the gap is
already very large and the temperature does not really change
this. The bump in the gap size seen at high temperatures can be
understood from equations \eqref{z} and \eqref{Deltaphi1}. The
temperature dependent part of equation \eqref{Deltaphi1} shows that
$\Delta\phi_g$ and thus $e\phi_e(R_C)$ will be constant for low
temperatures, $T\ll \mu_e$ $(r \gg R_C)$, and smaller for higher temperatures. 
Since
$\sinh^{-1}x\sim x$ for $\lvert x \rvert< 1$ and $\sinh^{-1}x\sim \log
2x$ for $x>1$ this can produce the bump. Physically the
thermal electrons in the crust means that $\mu_e$ does not have to be
as large to reach a certain density, which moves the crust boundary
further out, while the thermal electrons in the gap screens the
strange star surface charge and reduces the gap.

\section{Numerical solutions}
\subsection{Numerical procedure and boundary conditions}
To solve equations \eqref{mue} and \eqref{muN} numerically
we employ a shooting method which
integrates the equations from the quark core to the point of charge
neutrality in the crust using a standard Runge-Kutta routine. We then
vary ${d\mu_e}/{dr}$ at the starting point until equation
\eqref{diffsum} is fulfilled at charge neutrality so the crust remains
charge neutral beyond this point. Equations \eqref{mueff} would seem
to produce many more free parameters at the starting point than just
${d\mu_e}/{dr}$, but since $e\phi_e(r)$ at the quark core depends
on unknown details of the quark matter equation of state we keep it
free and explore the solutions for a wide range of this
parameter. Furthermore it was shown in the preceding sections that
the quantity $\Delta\phi_g=\phi_g(R)-\phi_g(R_C)$ is related to the
density of the crust in the charge neutral bulk through equations
\eqref{deltaphi} and \eqref{Deltaphi1} -- and as it turns out these
relations are fulfilled for the numerical solutions as well. Since
${d\mu_N}/{dr}$ may be determined at the starting point by
equation \eqref{diffsum} this leaves ${d\mu_e}/{dr}$ as the only
parameter to be determined. With this method we can thus choose
$e\phi_e(r)$ at the starting point (this corresponds to choosing
parameters for the equation of state for
the strange matter core) and a crust -- described by its density
($\Delta\phi_g$) and temperature at charge neutrality -- and find the
solution to equations \eqref{mue} and \eqref{muN} describing the transition between
the core and crust.

We have so far not specified the exact location of the starting point for the
integration because the discussion above does not depend
on this, and because the assumption of a color-flavor locked quark
core places the starting point at the core surface, whereas for a non
color-flavor locked core it should be taken in the charge neutral bulk
of the core. Color-flavor locked strange quark matter is charge
neutral in bulk because the pairing minimizes the energy for equal
quark Fermi momenta leading to equal numbers of u-, d-, and
s-quarks. However as discussed in \cite{Madsen:2001fu} when surface
effects are taken into account there will be a deficit of s-quarks
near the surface producing an overall positive surface charge. As
shown in \cite{Usov:2004iz} this gives a Coulomb barrier at the
surface of the order of $e\phi_e(R_s)\sim 36 \text{ MeV}$. For
color-flavor locked stars we thus take the starting point at the
surface of the quark core and find solutions for different values of
$e\phi_e(R_s)$ and different crusts.

In non color-flavor locked stars the deficit of s-quarks is global and
compensated by electrons in the bulk. The surface charge here arises
because some electrons near the surface leave the core and create an
atmosphere outside the quark phase. We therefore take the starting
point ``deep'' in the core (1000 fm below the surface; our results are
insensitive to this choice) where we
assume charge neutrality, $n_q^+(\mu_q^+)=n_e(\mu_e,T)$,
\begin{align}
\frac{\mu_q^{+^3}}{3\pi^2}=\frac{\mu_e^3}{3\pi^2}+\frac{\mu_eT^2}{3}.
\label{quarkcharge}
\end{align}
The chemical potential for the quark charge, $\mu_q^+$, defined by this
relation, was explored in
eg. \cite{Alcock:1986hz,Kettner:1994zs,Fu:2003} and found to be of the
order of $\mu_q^+\simeq 25 \text{ MeV}$. However the same surface
effect which gives rise to the strange quark deficit in color-flavor
locked stars will reduce the number of strange quarks further. We do not
explicitly include this effect in our calculations.

\subsection{Chemical composition of the crust}
The chemical composition of the crust depends on its origin and the
accretion history of the strange star. If the crust was created in the
supernova along with the strange star and has not changed since, it
will consist of cold catalyzed matter as described by the BPS equation
of state in \cite{BPS}. If the crust was accreted from a companion
star temperatures do not become high enough for the matter to reach
the equilibrium described by the BPS equation of state. Instead
hydrogen is burned at the surface to helium which is in turn burned
explosively in X-ray bursts and leaves a layer of heavy ashes at
densities exceeding $10^8 \text{ g/cm}^3$. The ashes then sink under
the weight of accreted matter and under the increasing pressure its
composition changes in a series of electron captures until neutron
drip sets in. This process has been investigated by Haensel \& Zdunik,
and the chemical composition as a function of density was given in
\cite{HZ1990}, assuming that the ashes left by the X-ray bursts are
composed of $^{56}$Fe. The composition of the X-ray burst ashes is a
matter of some debate however, and as it was shown by Schatz et al.\ in
\cite{Schatz:2001} that $rp$-processes in the X-ray bursts may lead to a
composition dominated by much heavier nuclei with $A\sim106$, Haensel
\& Zdunik recently revised the resulting chemical composition in
\cite{HZ2003}. However it was also suggested by Schatz et
al.~\cite{Schatz:2003} that the X-ray burst ashes may burn explosively
at densities around $\sim 10^9 \text{ g/cm}^3$ powering X-ray
superbursts in which case photodisintegration will lead to a
composition similar to that found in \cite{HZ1990}.

For our purpose we chose the relevant nucleus from these articles
based on the density in the crust at the point of charge neutrality --
the transition from gap to crust takes place over just a few fm so
there is no reason to assume any other nucleus will play a role. As
discussed the actual composition is rather uncertain, so we include
calculations for each of the three scenarios. The charge to
mass ratios are very similar, so the only significant difference turns
out to be related to the different values for the neutron drip density.
The chemical compositions used are shown at densities below
neutron drip in Table~\ref{tab1}. We hereafter refer to the
compositions in \cite{BPS,HZ1990,HZ2003} as BPS, HZ1990 and HZ2003
respectively. For the HZ1990 and HZ2003 compositions
which arise from explosive
helium burning we assume a composition of pure $^4$He below
$10^8~\text{g/cm}^3$. This transition should be more smooth, but it
hardly matters as the gap width is already very large at these
densities and small corrections from the composition will not affect
this.
\begin{table}[htbp]
\centering \footnotesize
\begin{tabular}{|c|c|c|c|c|c|}
\hline
\multicolumn{2}{|c|}{Ref. \cite{BPS} --
  BPS}&\multicolumn{2}{|c|}{Ref. \cite{HZ1990} --
  HZ1990}&\multicolumn{2}{|c|}{Ref. \cite{HZ2003} -- HZ2003}\\
\hline
$\rho_{max}$& Nucleus & $\rho_{max}$ &
Nucleus & $\rho_{max}$ & Nucleus \\
$[\text{g/cm}^3]$&&$[\text{g/cm}^3]$&&$[\text{g/cm}^3]$&\\
\hline
$8.1\times 10^6$ &  $^{56}$Fe &$1.494\times10^{9}$
&$^{56}$Fe&$3.517\times10^8$&$^{106}$Pd\\
$2.7\times 10^8$&$^{62}$Ni&$1.115\times10^{10}$&$^{56}$Cr&$5.621\times10^9$&$^{106}$Ru\\
$1.2\times 10^9$&$^{64}$Ni&$7.848\times10^{10}$ &$^{56}$Ti&$2.413\times10^{10}$&$^{106}$Mo\\
$8.2\times 10^9$&$^{84}$Se&$2.496\times10^{11}$ &$^{56}$Ca&$6.639\times10^{10}$&$^{106}$Zr\\
$2.2\times 10^{10}$&$^{82}$Ge&$6.110\times10^{11}$ &$^{56}$Ar&$1.455\times10^{11}$&$^{106}$Sr\\
$4.8\times 10^{10}$&$^{80}$Zn&&&$2.774\times10^{11}$&$^{106}$Kr\\
$1.6\times 10^{11}$&$^{78}$Ni&&&$4.811\times10^{11}$&$^{106}$Se\\
$1.8\times 10^{11}$&$^{76}$Fe&&&$7.785\times10^{11}$&$^{106}$Ge\\
$1.9\times 10^{11}$&$^{124}$Mo&&&&\\
$2.7\times 10^{11}$&$^{122}$Zr&&&&\\
$3.7\times 10^{11}$&$^{120}$Sr&&&&\\
$4.3\times 10^{11}$&$^{118}$Kr&&&&\\
\hline
\end{tabular}
\caption{Chemical composition of the crust as a function of density
  for the three scenarios discussed in the text. $\rho_{max}$ is the
  maximum density at which a nucleus is present before it undergoes
  electron capture.}
\label{tab1}
\end{table} 
\normalsize
\section{Results for color-flavor locked strange star cores}
\begin{figure}[!htbp]
\centering
\includegraphics[width=8.5cm]{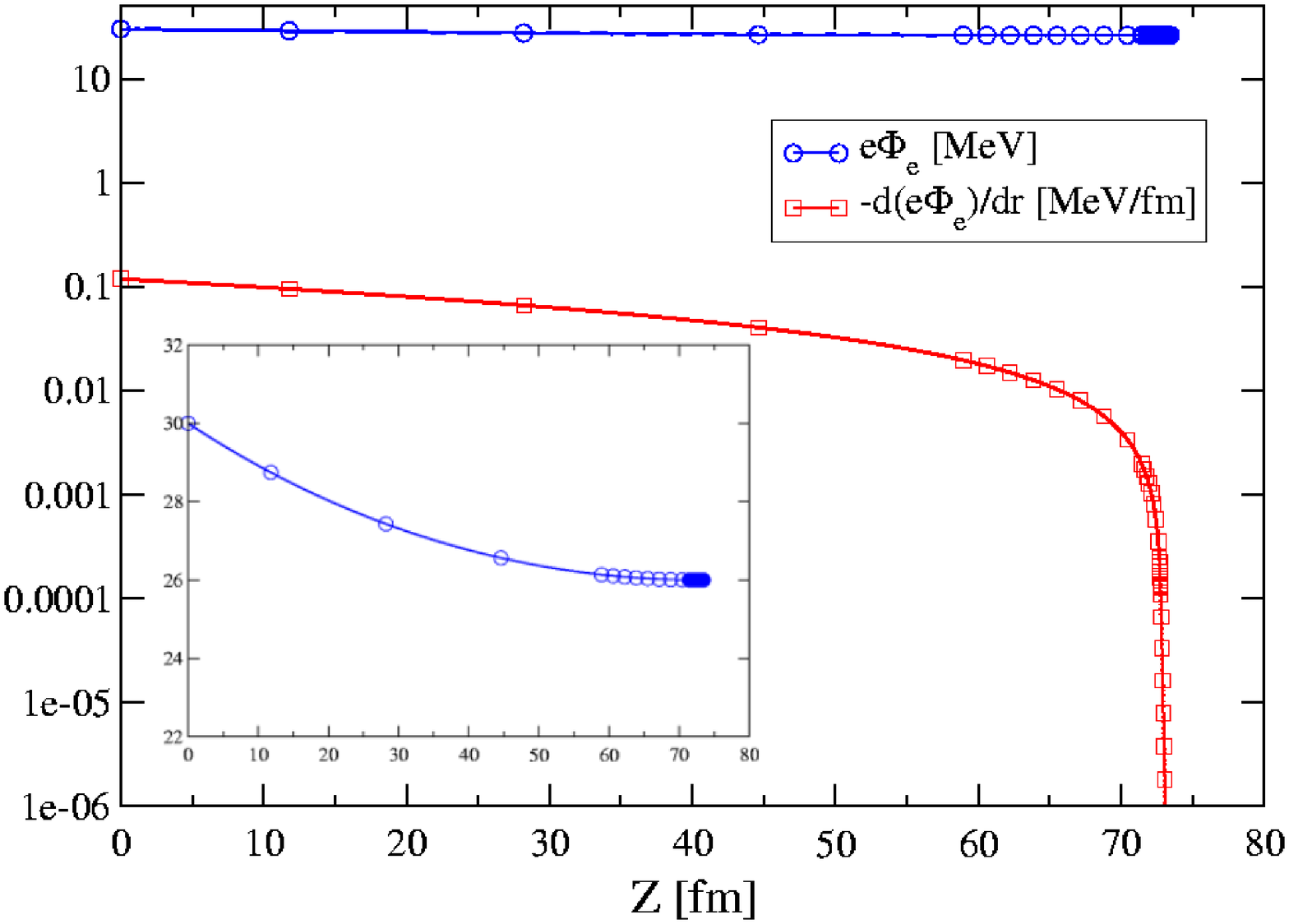}
\caption{Electric potential and field for a typical
  solution. Note how the field dies out in the charge neutral bulk of
  the crust. The inset focuses on the electric
  potential.}
\label{CFLfieldstructure}
\end{figure}

\begin{figure}[!htbp]
\centering
\includegraphics[width=8.5cm]{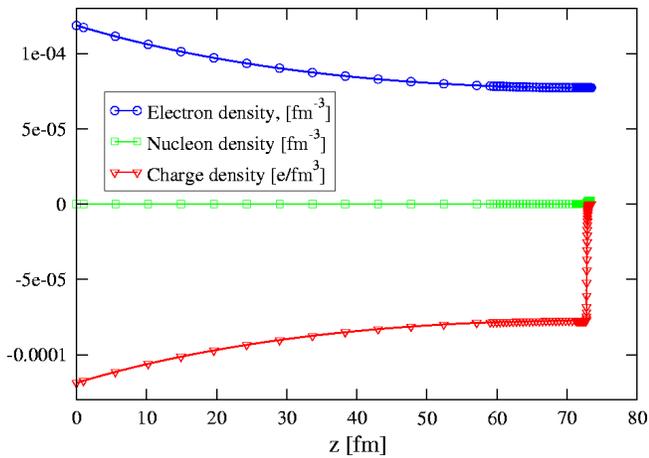}
\caption{Electron, nucleon and charge density for the solution 
  in Fig.~\ref{CFLfieldstructure}. The
  transition to the crust is shown in Fig.~\ref{CFLsmallstructure}.}
\label{CFLstructure}
\end{figure}

\begin{figure}[!htbp]
\centering
\includegraphics[width=8.5cm]{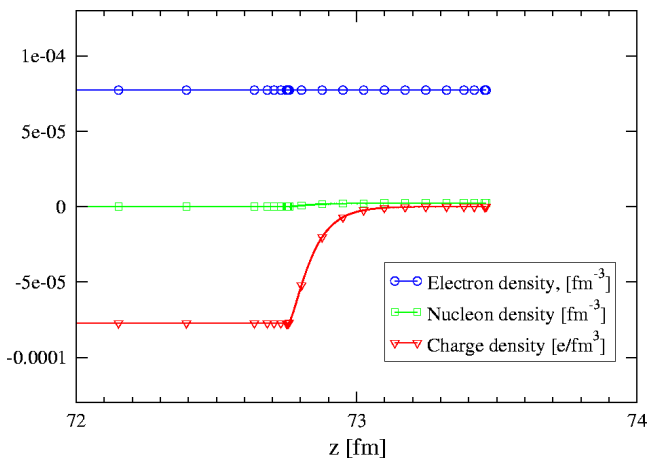}
\caption{Electron, nucleon and charge density during the
  transition to the crust for the solution in
  Figs.~\ref{CFLfieldstructure} and \ref{CFLstructure}. Nucleons appear
  over a few fm and charge neutrality is attained in the
  crust.}
\label{CFLsmallstructure}
\end{figure}

We have found solutions for a range of densities, compositions and
temperatures of the crust as well as for different electric
potentials at the quark core surface. The basic features of one such
solution with $\rho_{crust}=4\times 10^{11}\text{ g/cm}^3$, $T=0$,
$e\phi(R_s)=30\text{ MeV}$ and the crust composition in HZ2003 are
shown in Figs.~\ref{CFLfieldstructure}, \ref{CFLstructure}, and
\ref{CFLsmallstructure}. The
numerical solution conforms with the general behavior discussed
previously, but it should be noted that the actual transition to the crust
takes place over less than a fm; smaller than the radius of the relevant
nuclei. A model based on statistical
physics should hardly be trusted on such scales and the most
reasonable conclusion would probably be that the density is
essentially discontinuous at the crust boundary.

\begin{figure}[!htbp]
\centering
\includegraphics[width=8.5cm]{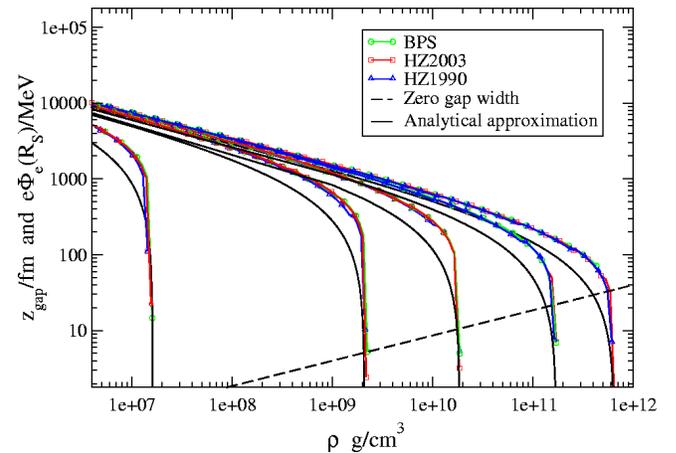}
\caption{Gap width dependence on density, crust composition and
  electric potential at the quark surface for color-flavor locked
  cores. From left to right $e\phi_e(R_S)$ is 1, 5, 10, 20 and 30 MeV.
  Curves for different composition are almost indistinguishable.
  The analytical approximations are based on Eq.~\eqref{approx} using the
  crust composition from the BPS equation of state. The dashed line
  shows the solution of Eq.~\eqref{approx} with $z_{gap}=0$ -- i.e. the
  density at which the gap width goes to zero for a specific
  $e\phi_e(R_S)$.}
\label{CFLrho}
\end{figure}

The dependence of the gap width on density, composition and electric
potential at the quark surface is explored at zero temperature in
Fig.~\ref{CFLrho} and compared to the analytical approximation in
Eq.~\eqref{approx}. In each case we plot (close) to the highest
possible density, which is either the neutron drip density given in
Table \ref{tab1} or the density at which $z_{gap}\sim 0$. The gap width is
insensitive to the choice of chemical composition, but it may be noted
that only for high $e\phi_e(R_s)$ is it possible to reach neutron drip
density before the gap width goes to zero. In particular the BPS and
HZ1990 compositions are limited by neutron drip at $e\phi_e(R_S)=30$
MeV whereas the gap width goes to zero before neutron drip for the
HZ2003 composition. As noted previously this choice of $e\phi_e(R_S)$
is therefore quite illustrative and we will often use it below.\\

\begin{figure}[!htbp]
\centering
\includegraphics[width=8.5cm]{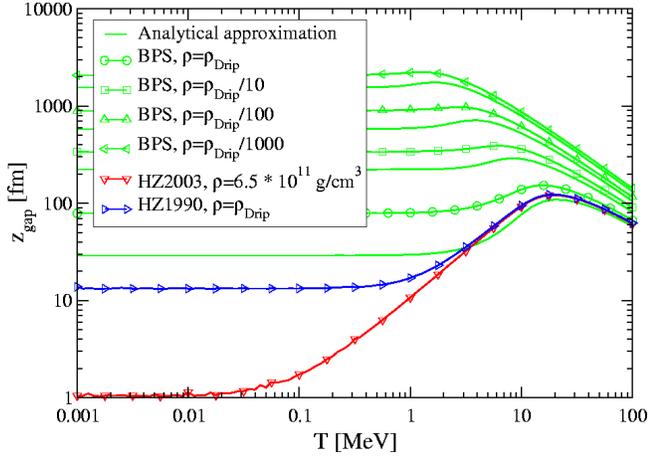}
\caption{Gap width dependence on temperature, density and crust
  composition for color-flavor locked cores with $e\phi_e(R_S)=30$
  MeV. Densities increase from top to bottom and curves without points
  show analytical approximations based on Eq.~\eqref{Deltaphi1}. For
  simplicity we only vary the composition for the maximum density
  crusts. The HZ2003 curve has a very low gap width because its
  density is not limited by neutron drip.}
\label{artikeltemp}
\end{figure}

Figure \ref{artikeltemp} shows the temperature dependence of $z_{gap}$
with $e\phi_e(R_S)=30$ MeV. Temperatures up to 100
MeV are shown to illustrate the full range of solutions, but one
should keep in mind that the Coulomb barrier is only 30 MeV or less,
and so can not hold the nuclei at these
temperatures. $e\phi_e(R_s)$ is set very high to allow high densities,
and for simplicity we only vary the composition for the maximum
density curves discussed above.  We recover the qualitative features
of the analytical approximation.
Except perhaps for the very early stages after formation in a supernova
explosion crust temperatures of isolated strange stars are expected to
be much smaller than $0.1-1$~MeV where the temperature effects for the
gap width become noticeable. Interesting temperatures may be reached in 
accreting binary systems to which we return in a later section. It may
also be relevant that the increase in gap width allows a strange
star to sustain a crust shortly after its formation in the supernova
even when temperatures are very high -- whether a crust would actually
form at that stage is a different matter.

\begin{figure}[!htbp]
\centering
\includegraphics[width=8.5cm]{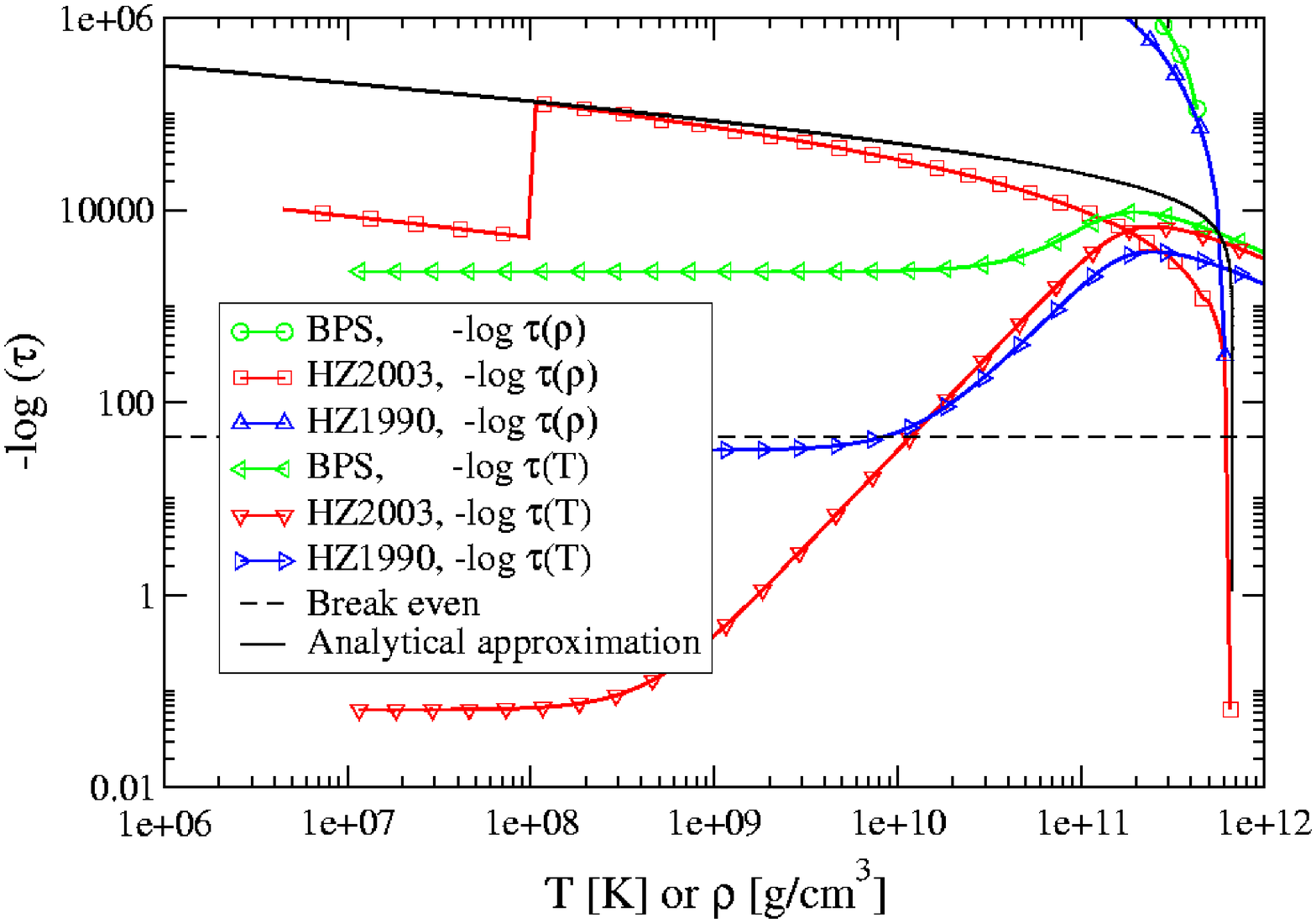}
\caption{Dependence of the transmission coefficient on density at
  $T=0$, and on temperature at maximum density with
  $e\phi_e(R_S)=30$ MeV. The kink in the HZ2003 density dependence is
  at the assumed transition to a hydrogen crust. The horizontal line
  marks the break even estimate for accreting stars in the
  text.}
\label{artikeltrans}
\end{figure}

Perhaps a better measure of the stability of the crust is the
transmission coefficient for ions through the gap, $\tau$, 
which following Alcock et al.\ 
\cite{Alcock:1986hz}, can be found in the WKB approximation by
\begin{align}
\tau=\exp{\left[-2\int_{z=0}^{z=z_g} \lvert k\rvert dz\right]}
\label{tau}
\end{align}
where $k=(\mu_N^2-m_N^2)^{1/2}$ is the wave number, and the
chemical potentials are known from the numerical solutions.
A few examples are shown in Fig.~\ref{artikeltrans}. Here we again choose
$e\phi_e(R_S)=30$ MeV and plot the dependence on density at zero
temperature, and the dependence on temperature for the maximum density
crust of each composition.

The integral in Eq.~\eqref{tau} can be approximated at $T=0$ by 
\begin{align}
\log{\tau}&\simeq-2\sqrt{2}m_N\int_0^{z_g}\left(\frac{Z}{m_N}e\phi_e(r)-\Delta\phi_g\right)^{1/2}dz\\
&=-2\sqrt{2}CZ\int_{x(R_S)}^{x(R_C)} x^{-3}(x^2-\Delta\phi_g)^{1/2}dx,
\end{align}
where
\begin{equation}
  x=\left(\frac{Z}{m_N}e\phi_e(r)\right)^{1/2},
\end{equation}
using Eq.~\eqref{mueff} for $\mu_N$ and Eq.~\eqref{coldpotential} for
$e\phi_e(r)$. Remembering that
$e\phi_e(R_C)=\frac{m_N}{Z}\Delta\phi_g$ this gives
\begin{equation}\label{tauapprox}
\begin{split}
&\log(\tau)=\\&-\frac{2\sqrt{2}CZ}{\Delta\phi_g^{1/2}}
\left[\cos^{-1}{\left(\frac{m_N\Delta\phi_g}{Ze\phi_e(R_S)}\right)^{1/2}}
-\left(\frac{1-\frac{m_N\Delta\phi_g}{Ze\phi_e(R_S)}}
{\frac{Ze\phi_e(R_S)}{m_N\Delta\phi_g}}\right)^{1/2}\right],
\end{split}
\end{equation}
where $\Delta\phi_g$ can be approximated by Eq.~\eqref{deltaphi}.
As seen from Fig.~\ref{artikeltrans} Eq.~\eqref{tauapprox} 
is a very reasonable approximation.
Assuming the HZ2003 crust composition at drip density, the low density
limit of Eq.~\eqref{tauapprox} is given as
\begin{equation}
\log(\tau) \approx -2.5\times 10^{4}\left[\frac{\pi}{2}\left(
\frac{\rho}{\rho_D}\right)^{-1/6}-1.84\right].
\end{equation}

The transmission coefficient may be better appreciated by estimating its
value when transmission of ions into the core through the barrier breaks
even with typical accretion rates in binary systems. That is, we
require the balance:
\begin{align}
\frac{\Dot{M}}{m_N}\leq N_{ion}\times f\times \tau\;,
\end{align}    
where $\Dot{M}$ is the mass accretion rate onto the strange star from a
companion, $N_{ion}$ is the number of
ions, whose motion about their lattice position allows them to strike
the barrier, and $f$ is the oscillation frequency of this
motion. $N_{ion}$ may be taken as roughly the number of ions within
one lattice distance, $a\sim 200$ fm, from the crust boundary, and
following again Alcock et al.\ \cite{Alcock:1986hz}, the oscillation frequency
should be less than 1 MeV. For a strange star accreting $10^{-10}
\text{M}_\odot\text{yr}^{-1}$ we then arrive at the condition
\begin{equation}
\begin{split}
-\log \tau \geq
 44.6-\log\left(\frac{\dot{M}}{10^{-10}\text{M}_\odot\text{yr}^{-1}}\right) +\log\left(\frac{f}{\text{MeV}}\right) \\+\log\left(\frac{\rho}{4.3\times 10^{11}\text{ g/cm}^3}\right)+\log\left(\frac{a}{200 \text{ fm}}\right)+2\log\left(\frac{R_S}{10 \text{ km}}\right) \label{stabtau}
\end{split}
\end{equation}
as indicated in Fig.~\ref{artikeltrans} (all logarithms are base $e$). 
It may then be noted that the
maximum density models shown in Fig.~\ref{artikeltrans} -- except the one
using the BPS chemical composition -- will interact with the core at
low temperature and are not stable at these densities. Only the BPS
equation of state is thus able to remain stable at neutron drip
density even with $e\phi_e(R_S)=30$ MeV. For other compositions and
lower $e\phi_e(R_S)$ stable crust densities are lower depending on the
temperature, see Fig.~\ref{stabrho}.

\section{Results for Non color-flavor locked strange star cores}
\begin{figure}[!htbp]
\centering
\includegraphics[width=8.5cm]{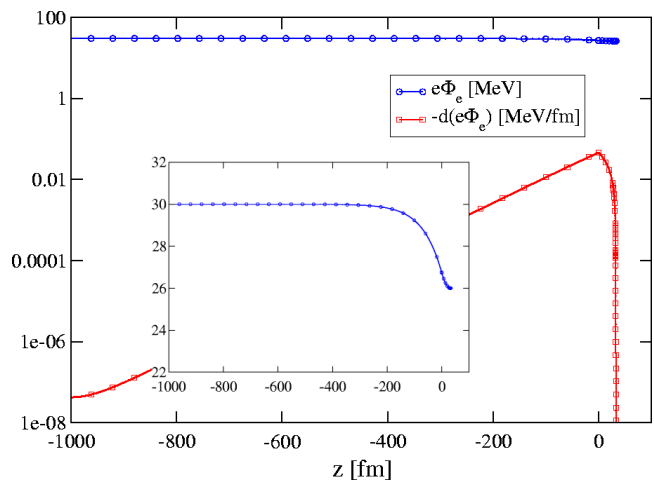}
\caption{Electric potential and field for a solution in
  the non color-flavor locked case. The inset focuses on the electrical
  potential.}
\label{NCFLfieldstructure}
\end{figure}

\begin{figure}[!htbp]
\centering
\includegraphics[width=8.5cm]{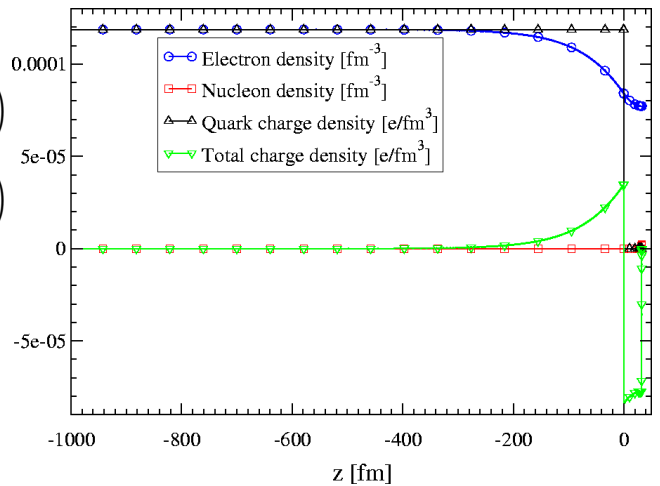}
\caption{Electron, nucleon, quark charge density and
  total charge density for the same solution as in
  Fig.~\ref{NCFLfieldstructure}. The transition to the crust is
  shown in Fig.~\ref{NCFLsmallstructure}.}
\label{NCFLstructure}
\end{figure}

\begin{figure}[!htbp]
\centering
\includegraphics[width=8.5cm]{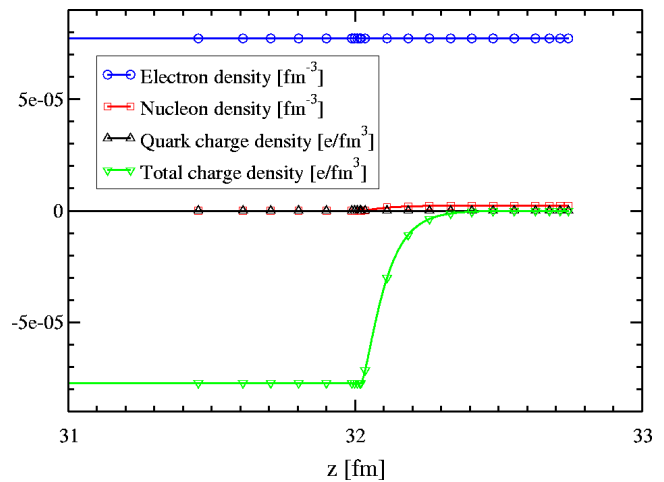}
\caption{Electron, nucleon, quark charge and
  total charge density during the transition from gap to crust for the
  same solution as in Figs. \ref{NCFLfieldstructure} and
  \ref{NCFLstructure}.}
\label{NCFLsmallstructure}
\end{figure}

We have performed similar calculations for the case of non
color-flavor locked cores (``ordinary'' strange stars), 
and the structure of a solution with
$\rho_{crust}=4\times 10^{11}\text{ g/cm}^3$, $T=0$, $\mu_q^+=30\text{
MeV}$ and the HZ2003 crust composition is shown in
Figs.~\ref{NCFLfieldstructure}, \ref{NCFLstructure}, and
\ref{NCFLsmallstructure}. Charge
neutrality is imposed at the starting point 1000 fm into the core and as
we approach the surface electrons become scarcer and consequently the
charge density and electric field rises. At the surface the quarks and
their positive charge vanish, the charge density becomes negative and
the field starts decreasing. The solution from the surface out looks
like the color flavor-locked case with $e\phi_e(R_s)=\mu_e(R_s)$.

\begin{figure}[!htbp]
\centering
\includegraphics[width=8.5cm]{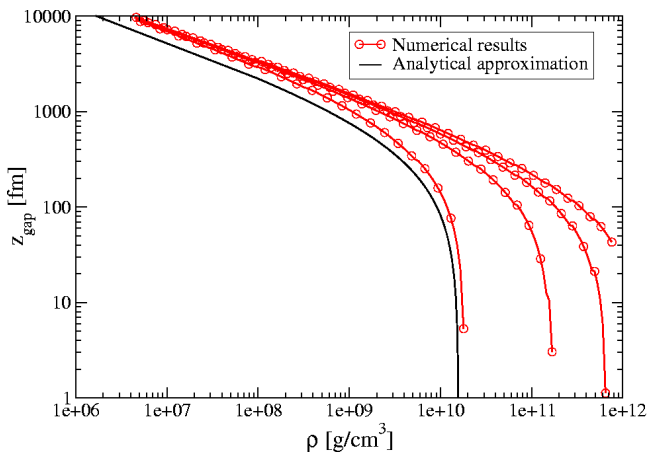}
\caption{Dependence of the gap width on density using the HZ2003 crust
  composition. From left to right $\mu_q^+=10,\;20,\; 30,\; 40 \text{
  MeV}$. The analytical approximation explained in the text assumes
  $\mu_q^+=10 \text{ MeV}$.}
\label{NCFLrho}
\end{figure}

We again calculate the dependence of the gap width on density shown in
Fig. \ref{NCFLrho} for $T=0$ and different choices of
$\mu_q^+=10,\;20,\;30,\;40 \text{ MeV}$ from left to right. We only
show models using the HZ2003 crust composition, since the gap width is
in this case so insensitive to the choice of composition that the
curves would be indistinguishable. Note again that only for the
highest choice of $\mu_q^+= 40\text { MeV}$ does the crust reach
neutron drip density before the gap goes to zero.
The gap width is approximated by Eq.~\eqref{approx} using
$e\phi_e(R_S)=\mu_q^+=10~\text{MeV}$. For a bare strange star we would have
$e\phi_e(R_S)=3\mu_q^+/4$ \cite{Alcock:1986hz}, whereas for dense
crusts $e\phi_e(R_S)\simeq \mu_q^+$, but since large gap widths are
insensitive to $e\phi_e(R_S)$ we use $e\phi_e(R_S)=\mu_q^+$ in all
analytical approximations.

\begin{figure}[!htbp]
\centering
\includegraphics[width=8.5cm]{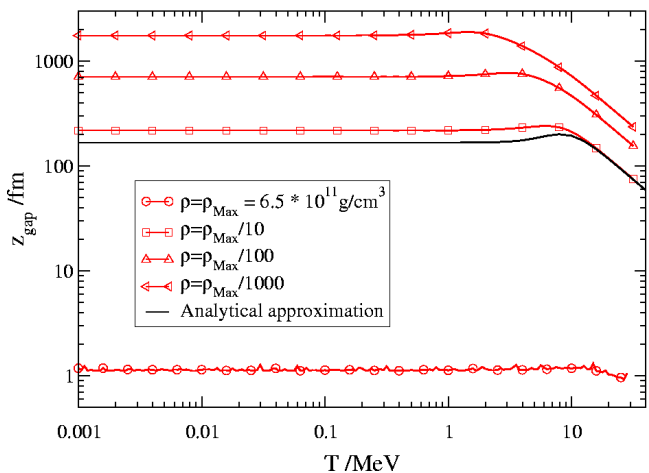}
\caption{Gap width dependence on temperature and density
  for non color-flavor locked cores with $\mu_q^+=30$ MeV. The maximum
  density is taken to be the density at which $z_{gap}\simeq 1$ fm as
  this is as close to zero gap width as is numerically reasonable -- note
  that variations in the numerical results are significant only at this
  scale. The analytical estimate assumes $\rho=\rho_{max}/10$.}
\label{NCFLtemp}
\end{figure}

The temperature dependence of the gap width is explored in
Fig.~\ref{NCFLtemp} for the case of $\mu_q^+= 30\text{ MeV}$
with the HZ2003 crust
composition. We note that the gap width declines steeply at $T\sim
10\text{ MeV}$ and reaches zero at $T\sim30 \text{ MeV}$  in contrast
to the color-flavor locked case. This behavior is caused by the
condition of charge neutrality at the starting point in the core
(Eq.~\eqref{quarkcharge}),
which reduces $\mu_e$ significantly at temperatures comparable to
$\mu_q^+$. Other that that we note again the bump in the gap width
caused by the condition of charge neutrality in the crust. The lowering
of $\mu_e$ entirely suppresses the bump at high densities. 
The gap width is approximated in Fig.~\ref{NCFLtemp} by Eq.~\eqref{z}
using $e\phi_e(R_S)=\mu_e(z=-1000 \text{ fm})$ found from
Eq.~\eqref{quarkcharge}. $e\phi_e(R_C)$ is estimated as before from
Eq.~\eqref{sum} and \eqref{Deltaphi1}.

\begin{figure}[!htbp]
\centering
\includegraphics[width=8.5cm]{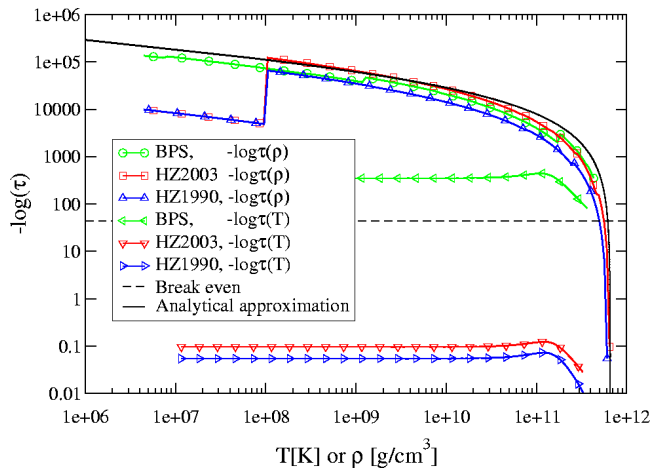}
\caption{Dependence of the transmission coefficient on
  density at zero temperature and temperature at maximum density for
  non color-flavor locked strange star cores with $\mu_q^+=30\text{
  MeV}$. The kink in the HZ1990 and HZ2003 dependence corresponds to
  the assumed transition to a hydrogen crust. The break even estimate
  is explained in the text and the analytical approximation is based
  on Eq.~\eqref{tauapprox} with the HZ2003 crust composition at
  drip density and $e\phi_e(R_S)=\mu_q^+$.}
\label{NCFLtrans}
\end{figure}

The variation of transmission coefficient with density and temperature
is shown in Fig.~\ref{NCFLtrans}. We again note that the bump is
suppressed compared to the color-flavor locked case and that only
crusts with the BPS composition can reach neutron drip before they
become unstable against transmission through the Coulomb barrier.

\section{Discussion}
We have expanded the treatment of gaps below strange star crusts to
include effects of pressure and gravity on the crust nuclei and found
the structure of the transition from gap to crust. Overall our results
are as one would expect with a very sharp transition from gap to
crust, gap widths ranging from a few fm at densities near neutron
drip over a few thousand fm at densities close to the maximum to
$10^{10}$ fm for very thin crusts. Crust densities are consequently
limited below neutron drip unless one assumes an extreme Coulomb
barrier height. If the gap width must be at least a crust lattice distance
($\sim 200$ fm) and a reasonable value of $e\phi_e(R_S)\simeq 20 \text{MeV}$
(or $e\phi_e(z=-1000\text{ fm})\simeq 20\text{ MeV}$) is assumed then the
crust density is limited to a few times $10^{10} \text{ g/cm}^3$.

\begin{figure}[!htbp]
\centering
\includegraphics[width=8.5cm]{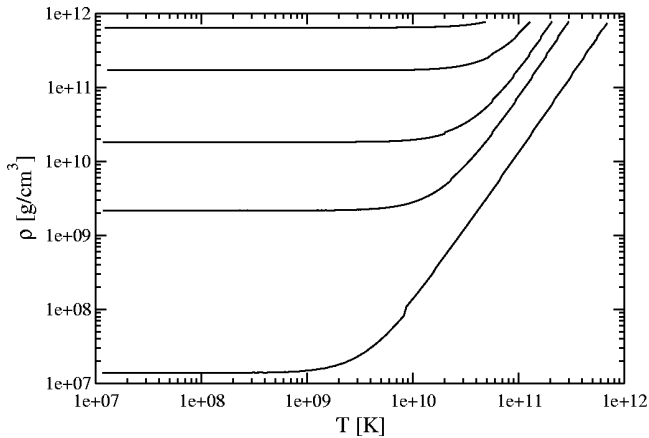}
\caption{Highest stable density according to
  Eq.~\eqref{stabtau} at given temperature and $e\phi_e(R_S)=
  1,5,10,20,30$~MeV from below. The
  crust composition is from HZ2003 and we stop at the temperature
  corresponding to neutron drip density.}
\label{stabrho}
\end{figure}

The variation of the gap width with temperature is noteworthy in that
it initially increases with temperature which is a qualitatively new
feature. However this increase only takes place at temperatures below
1 MeV ($10^{10}$ K) for crust densities which give very small gap
widths of the order of 1 fm such as in the case of the HZ2003 crust at
maximum density in Figs.~\ref{artikeltemp} and \ref{artikeltrans}. A
realistic crust should have a sufficient gap width to be stable
against strong interaction with the core, and examples of such crusts could
be the HZ1990 and BPS crusts in Figs.~\ref{artikeltemp} and
\ref{artikeltrans} -- the HZ1990 crust is almost stable, and lowering
the density a little to make it so would not change the dependence on
temperature much. If we apply the criterion in Eq.~\eqref{stabtau} for
crust stability we can find the highest stable density at given
temperature and Coulomb barrier height, $e\phi_e(R_S)$. Numerical
results for this are shown in Fig.~\ref{stabrho} and we see that
temperature effects are then not important below $10^{10}$ K.
If we demand at least a
crust lattice distance in non accreting systems the crust becomes even
more stable against variations in the temperature.

The thermal structure of accreting strange stars was investigated by
Miralda-Escud\'e et al.\ \cite{Miralda:1990} who found temperatures
around $10^7$ K at the crust boundary for strange stars accreting
$10^{-11}$ to $10^{-9.83} \text{ M}_\odot \text{ yr}^{-1}$ assuming that
10 MeV is released per nucleon converted to strange quark matter (bag
constant $B=103 \text{ MeV/fm}^3$). In this case the gap width would
not be affected at all, and one might as well use the zero temperature
results. Superbursters however are known to have much higher accretion
rates around 0.1-0.3 times the Eddington accretion rate,
$M_{\text{Edd}}\sim 10^{-8}\text{ M}_\odot\text{ yr}^{-1}$, and in a
recent investigation Page and Cumming \cite{Page:2005} found
temperatures for such stars in the range $10^8-10^9$ K at the crust
boundary. This work was motivated by the difficulties in achieving
carbon ignition at observed column depths in superburst models on
neutron stars -- see Cumming et al. 2005 \cite{Cumming:2005} for a
recent review -- and it was found that superbursts may ignite at the
right column density on strange stars provided that neutrino emission
in the core is slower than direct Urca which implies that strange
quark matter should be a color superconductor. During the superburst
itself temperatures in the range $2-7 \times 10^9$ K are reached for a
few hours \cite{Cumming:2001, Schatz:2003}. This is again too low for
significant temperature effects and for stable crusts one may as well
use the zero temperature expressions.

The present treatment could be improved by including general
relativity instead of Newtonian gravity, and by using a less simplistic
model for the equation of state for the crust. Such improvements are
necessary to calculate the properties of the whole star, but we expect
the changes in the gap properties to be minor compared to the results
presented here. The same can be said about a more detailed treatment of
the quark phase, which has here merely been treated as a source of a
given total mass and starting value for the electron chemical potential.
However, as pointed out in \cite{Jaikumar:2005}, the electric potential
in the outer layers of the strange core could be qualitatively different
from the behavior assumed here if a mixed phase of quark nuggets and
electrons can be realized.

Another important oversimplification involves the assumption of a constant
pairing gap all the way to the quark core surface in the treatment
of quark matter in the CFL-phase.
In fact the pairing gap $\Delta_{\text{CFL}}$ may be envisaged to vanish
within a surface distance of order 
$1/\Delta_{\text{CFL}}\approx {\text{2--20~fm}}$
for gap energies between 100 and 10~MeV. Possibly a
range of phases appear in the extreme uppermost layers, including the
possibility that the very surface resembles the ungapped ordinary quark
phase also discussed above. A detailed treatment of gapped phases near a
surface remains to be performed.

In conclusion the results of our expanded treatment, while
quantitatively somewhat different, are qualitatively
consistent with previous work in the literature with regards to
gap width, transmission coefficient and possible crust densities. The
increase in gap width with temperature is new but significant only at
very high temperatures not likely realized. Perhaps the most useful
results of the investigation are the various analytical approximations
which have been derived and shown to fit the full numerical solutions
very well, since these provide better physical insight and can be
easily used in models for strange star rotation, glitches, accretion and
instabilities.

\acknowledgments
This work was supported by the Danish Natural Science Research Council.
We thank Fridolin Weber for useful comments on an early draft.

\end{document}